%% file: main.tex
\documentclass[conference,9pt,letterpaper]{IEEEtran}

\usepackage{cite}
\usepackage{amsmath,amssymb}
\usepackage{booktabs}
\usepackage{multirow}
\usepackage[hidelinks]{hyperref}

\newif\ifanonymous
\anonymousfalse

\newcommand{\projectname}{DuetFHE}
\title{\projectname: Compiling Dual-Lane Nonlinear Circuits for Complex-Packed CKKS Inference}
\ifanonymous
  \author{\IEEEauthorblockN{Anonymous Authors}}
  \hypersetup{pdfauthor={}}
\else
  \input{authors}
\fi

\hypersetup{pdftitle={DuetFHE: Compiling Dual-Lane Nonlinear Circuits for Complex-Packed CKKS Inference},pdfsubject={},pdfkeywords={}}

\begin{document}
\maketitle
\pagestyle{empty}
\thispagestyle{empty}

\begin{abstract}
\input{sections/abstract}
\end{abstract}

\input{sections/introduction}
\input{sections/background}

\input{sections/method}
\input{sections/compiler}
\input{sections/evaluation}
\input{sections/related-work}
\input{sections/conclusion}

\bibliographystyle{IEEEtran}
\bibliography{references}
\end{document}

%% file: authors.tex
\newcommand{\arxivauthors}{Ke Changke, Jian Wang, Xuan Xu, Jiqiang Liu}

\author{
  \IEEEauthorblockN{\arxivauthors}
  \IEEEauthorblockA{Beijing Jiaotong University, Beijing, China\\
    \texttt{changke@bjtu.edu.cn}\\
    Corresponding author: Jiqiang Liu (\texttt{jqliu@bjtu.edu.cn})}
}

\hypersetup{pdfauthor={\arxivauthors}}

%% file: sections/abstract.tex
Complex packing in CKKS stores two real streams in the real and imaginary lanes of each slot.
Direct squaring couples the lanes, while squaring them separately requires two ciphertext--ciphertext multiplications (HMults).
DuetFHE is a GPU compiler that uses a phase transformation to square both lanes independently with one HMult.
It reuses dual-lane squares in polynomial activations and the modular-reduction stage of CKKS bootstrapping.
The compiler further plans ciphertext scales and reuses conjugates to sustain the complex-packed representation across stages.
On a single RTX 5090, DuetFHE accelerates four polynomial activations by $1.60$--$1.79\times$ over direct evaluation on two real ciphertexts.
For two-image AESPA ResNet-20 inference, it maintains packed execution across convolutions, square activations, and bootstrapping, achieving a $1.97\times$ end-to-end speedup over the fastest tested EasyFHE batched baseline on both the RTX 5090 and RTX PRO 6000.

%% file: sections/introduction.tex
\section{Introduction}
\label{sec:introduction}
Fully homomorphic encryption (FHE) enables servers to evaluate machine learning models on encrypted inputs.
Complex packing in CKKS~\cite{cheon2017ckks} stores two real streams in the real and imaginary lanes of each slot and lets them share linear operations~\cite{boemer2019ngraphhe2,park2025powerformer,xu2025multiimage}.

Directly evaluating a nonlinear polynomial on a packed ciphertext can couple the lanes.
For a slot $z=x+iy$, the desired lane-wise square is $x^2+iy^2$, whereas direct squaring gives $x^2-y^2+2ixy$.
Separating the lanes and squaring them independently requires two ciphertext--ciphertext multiplications (HMults), and subsequent packed operators require repacking the results.

The same bottleneck recurs in polynomial activations such as GELU and in the modular-reduction stage of CKKS bootstrapping (\textsf{EvalMod})~\cite{lee2021highprecision}.
Keeping the lanes packed across these circuits and surrounding layers requires compatible slot layouts, modulus levels, and scales.
Conjugation key switches and the memory footprint of intermediate ciphertexts also affect how much the arithmetic savings improve GPU inference.

We present \projectname{}, a GPU compiler for complex-packed CKKS inference.

Our main contributions are as follows:
\begin{itemize}
  \item \textbf{Dual-Lane Squaring via an Exact Phase Transformation.}
  Our construction squares two real streams independently with one HMult, compared with two HMults for two independent real squares.
  \item \textbf{Composable Shared-Square Circuits.}
  We reuse dual-lane squares in polynomial activations and throughout \textsf{EvalMod}, including its double-angle stages.
  \item \textbf{State-Aware GPU Compilation.}
  The compiler aligns ciphertext states, reuses conjugates, and streams compatible operators within the GPU memory budget.
\end{itemize}

On an RTX 5090, \projectname{} accelerates four polynomial activations by $1.60$--$1.79\times$ over direct polynomial evaluation on two real ciphertexts.
For two-image AESPA ResNet-20 inference, it achieves a $1.97\times$ end-to-end speedup over the fastest tested EasyFHE batched baseline on both the RTX 5090 and the RTX PRO 6000.
For BERT-base inference on the THOR workload, it achieves end-to-end speedups of $1.33\times$ on the RTX 5090 and $1.20\times$ on the RTX PRO 6000 over EasyFHE's THOR implementation.

%% file: sections/background.tex
\section{Background and Motivation}
\label{sec:background}

\subsection{CKKS Execution Costs on GPUs}
\label{sec:bg-ckks}

CKKS supports approximate SIMD arithmetic on vectors of up to $N/2$ complex slots in the polynomial ring $R_{Q_\ell}=\mathbb{Z}_{Q_\ell}[X]/(X^N+1)$, where $N$ is the ring dimension and $Q_\ell$ is the ciphertext modulus~\cite{cheon2017ckks}.
In the residue number system (RNS), the modulus at level $\ell$ is a product of active primes, $Q_\ell=\prod_{j=0}^{\ell}q_j$~\cite{cheon2019fullrns}.
An HMult operates across these primes and is typically followed by relinearization and rescaling.
Rotations and conjugation also require key switching, so the cost of a nonlinear circuit depends on these operations as well as its HMult count.

\subsection{Complex Packing and Independent Nonlinear Computation}
\label{sec:bg-complex}

Complex packing stores two real streams in the real and imaginary lanes of each CKKS slot, with $z=\mathcal{P}(x,y)=x+iy$.
For real plaintext weights $w$, the identity $w\mathcal{P}(x,y)=\mathcal{P}(wx,wy)$ allows both streams to share one plaintext--ciphertext multiplication.
A common slot rotation likewise acts independently on both lanes.

More generally, for $u=\mathcal{P}(u_1,u_2)$ and $v=\mathcal{P}(v_1,v_2)$, the general dual-lane product is defined by $\mathcal{D}_\times(u,v)=\mathcal{P}(u_1v_1,u_2v_2)$.
Its conventional construction is
\begin{equation}
    \mathcal{D}_\times(u,v)
    =\frac{(u+\bar{u})(v+\bar{v})-i(u-\bar{u})(v-\bar{v})}{4}.
    \label{eq:bg-dual-product}
\end{equation}
This construction uses two HMults even when $u=v$ and requires a conjugation key switch for each distinct input whose conjugate is unavailable.

Real-slot batching can also square two real streams with one HMult by placing their values in separate slots of a single ciphertext.
Its applicability depends on the available slot capacity and the slot layouts required by surrounding linear layers and bootstrapping stages.

%% file: sections/method.tex
\section{Composable Dual-Lane Computation}
\label{sec:method}

\subsection{Dual-Lane Squaring}
\label{sec:method-arithmetic}

For a packed slot $z=x+iy$, let $a=z+\bar z=2x$ and $b=z-\bar z=2iy$.
When the two operands are identical, the standard two-HMult construction in Eq.~\eqref{eq:bg-dual-product} can be factored into one product using a fixed ring phase.
In the CKKS ring $R_{Q_\ell}$, choose $M=X^{N/4}$.
On the slot representatives $\zeta_j=\omega^{5^j}$, where $\omega=e^{\pi i/N}$, this monomial satisfies $M(\zeta_j)^2=i$ because $5^j\equiv1\pmod 4$.
The dual-lane square is therefore
\begin{equation}
\label{eq:dual_lane_square}
\operatorname{DSq}(z)
  =\frac{(a+Mb)(a-Mb)}{4}
  =\frac{a^2-M^2b^2}{4}
  =x^2+iy^2.
\end{equation}
The product cancels the cross terms, and the phase returns the second square to the imaginary lane.
Multiplication by $M$ is an exact signed permutation of ring coefficients and consumes no modulus level.
The primitive uses one HMult and requires one conjugation key switch when $\bar z$ is unavailable.
The compiler folds the normalization factor into the effective output scale and tracks that scale at subsequent uses.

\subsection{Packed Polynomial Evaluation and Bootstrapping}
\label{sec:dual-poly}

Polynomial activations can share nonlinear intermediates across terms in each lane.
DuetFHE uses activation symmetry to expose squared arguments and reuses dual-lane squares when generating Chebyshev basis terms.
For higher degrees, it selects among polynomial decompositions by HMult count within the available modulus-level budget.
General dual-lane products handle combinations that require distinct operands.

In cosine-based \textsf{EvalMod}, DuetFHE evaluates the base polynomial lane-wise and carries the packed result through the successive double-angle stages.
Each double-angle stage advances both lanes with one HMult while preserving the stage count and algebraic multiplicative depth.
The packed output passes to the subsequent bootstrapping stages without extracting and repacking the lanes between stages.

%% file: sections/compiler.tex
\section{DuetFHE Compiler Design}
\label{sec:compiler}

\begin{figure*}[t]
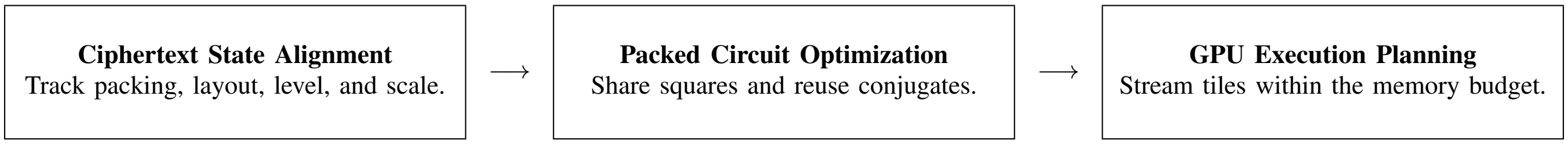

\centering
\begingroup
\setlength{\fboxsep}{5pt}
\small
\fbox{\parbox[c][3.3\baselineskip][c]{0.27\textwidth}{\centering
  \textbf{Ciphertext State Alignment}\\
  Track packing, layout, level, and scale.}}
\enspace $\longrightarrow$ \enspace
\fbox{\parbox[c][3.3\baselineskip][c]{0.27\textwidth}{\centering
  \textbf{Packed Circuit Optimization}\\
  Share squares and reuse conjugates.}}
\enspace $\longrightarrow$ \enspace
\fbox{\parbox[c][3.3\baselineskip][c]{0.27\textwidth}{\centering
  \textbf{GPU Execution Planning}\\
  Stream tiles within the memory budget.}}
\endgroup
\caption{The \projectname{} compilation pipeline.}
\label{fig:overview}
\end{figure*}

DuetFHE lowers model operators into a dual-lane intermediate representation (IR) and emits an execution plan for the EasyFHE GPU backend~\cite{easyfhe2026}.
Figure~\ref{fig:overview} shows how it keeps compatible operators packed through lowering, circuit optimization, and execution planning.

\paragraph{Ciphertext state alignment}
The IR records each value's packing, slot layout, modulus level, and effective scale.
At each nonlinear call site, the compiler selects a circuit compatible with the input state and the required output state.
It checks scale and level compatibility at branch merges and aligns shared values at the consumers that require a different state.
This allows several consumers to use the same nonlinear intermediate.

\paragraph{Conjugate reuse}
Dual-lane squares and products consume conjugated operands.
The compiler reuses available conjugates and derives them through compatible linear operations when possible.
For example, it can add the available conjugates of two inputs to obtain the conjugate of their sum without another key switch.
It emits a direct conjugation when the graph cannot supply one.

\paragraph{Memory-aware execution}
For compatible linear--nonlinear--linear chains, the compiler streams activation tiles into the following projection while preserving the required layout, level, and scale.
The schedule releases intermediates after their final use and reuses storage across nonoverlapping lifetimes.
It checks GPU memory against the combined footprint of live ciphertexts and resident resources.

%% file: sections/evaluation.tex
\section{Evaluation}
\label{sec:evaluation}

\begin{table*}[t]
\centering
\caption{Single-GPU inference. Speedups are relative to the first scheme for each workload on each GPU.}
\label{tab:e2e}
\footnotesize
\setlength{\tabcolsep}{3pt}
\renewcommand{\arraystretch}{1.12}
\begin{tabular*}{\textwidth}{@{\extracolsep{\fill}}llrrrrrrrr@{}}
\toprule
\multicolumn{2}{c}{CKKS: $(\log_2 N,\mathrm{dnum},\log_2(PQ))$} & \multicolumn{4}{c}{RTX 5090 (32\,GB)} & \multicolumn{4}{c}{RTX PRO 6000 Blackwell (96\,GB)}\\
\cmidrule(lr){3-6}\cmidrule(l){7-10}
\multicolumn{1}{@{}c}{Workload} & \multicolumn{1}{c}{Scheme}
 & \multicolumn{1}{c}{s/item} & \multicolumn{1}{c}{Items/s}
 & \multicolumn{1}{c}{\shortstack{Peak active\\(GiB)}} & \multicolumn{1}{c}{\shortstack{Output err.\\($10^{-6}$)}}
 & \multicolumn{1}{c}{s/item} & \multicolumn{1}{c}{Items/s}
 & \multicolumn{1}{c}{\shortstack{Peak active\\(GiB)}} & \multicolumn{1}{c@{}}{\shortstack{Output err.\\($10^{-6}$)}}\\
\midrule
\multirow{2}{*}{\shortstack[l]{BERT-base (THOR)\\$B=1;\ (16,3,2490)$}} & EasyFHE & 52.077 ($1.00\times$) & 0.0192 & 25.56 & $21319$ & 43.211 ($1.00\times$) & 0.0231 & \textbf{68.93} & $\boldsymbol{16683}$\\
 & \projectname{} & \textbf{39.256} ($\boldsymbol{1.33\times}$) & \textbf{0.0255} & \textbf{25.29} & $\boldsymbol{18334}$ & \textbf{35.973} ($\boldsymbol{1.20\times}$) & \textbf{0.0278} & 75.67 & $18596$\\
\midrule
\multirow{2}{*}{\shortstack[l]{BERT-base (MOAI)\\$(16,3,3438)$}} & EasyFHE & / & / & OOM & / & 16.847 ($1.00\times$) & 0.0594 & \textbf{67.18} & $352799$\\
 & \projectname{} & 16.030 & 0.0624 & 30.89 & $6070$ & \textbf{13.902} ($\boldsymbol{1.21\times}$) & \textbf{0.0719} & 73.67 & $\boldsymbol{7312}$\\
\midrule
\multirow{4}{*}{\shortstack[l]{AESPA ResNet-20\\$B=2;\ (16,3,2490)$}} & Independent real & 0.67193 ($1.00\times$) & 1.49 & \textbf{14.81} & $4.20$ & 0.67093 ($1.00\times$) & 1.49 & \textbf{14.81} & $4.31$\\
 & Batch, fused square & 0.64754 ($1.04\times$) & 1.54 & 14.99 & $4.28$ & 0.64492 ($1.04\times$) & 1.55 & 14.99 & $\boldsymbol{4.16}$\\
 & Batch, native square & 0.64073 ($1.05\times$) & 1.56 & 14.98 & $\boldsymbol{4.09}$ & 0.64485 ($1.04\times$) & 1.55 & 14.98 & $4.19$\\
 & \projectname{} & \textbf{0.32557} ($\boldsymbol{2.06\times}$) & \textbf{3.07} & 15.11 & $5.53$ & \textbf{0.32751} ($\boldsymbol{2.05\times}$) & \textbf{3.05} & 15.11 & $5.48$\\
\bottomrule
\end{tabular*}
\par\noindent\raisebox{-2pt}[\height][\depth]{\parbox{\textwidth}{\raggedright Items: sentence pairs (THOR), positions (MOAI; 256 per batch), or images (ResNet-20; two per batch).
A slash denotes an unavailable metric due to GPU OOM.}}
\end{table*}

\subsection{Experimental Setup}
\label{sec:eval-setup}

\textbf{Implementation.}
DuetFHE runs on EasyFHE's CUDA backend~\cite{easyfhe2026} with 64-bit RNS arithmetic and single-prime rescaling.
The packed polynomial activation circuits use the standard complex representation in Sec.~\ref{sec:method-arithmetic}.
The packed ResNet-20, \textsf{EvalMod}, and full bootstrapping circuits use a scaled complex representation.

\textbf{Hardware.}
The evaluation targets two NVIDIA Blackwell GPUs: the GeForce RTX 5090 (32\,GB) and the RTX PRO 6000 Blackwell (96\,GB).
Each run uses one GPU.

\textbf{Benchmarks and Baselines.}

We benchmark ResNet-20 with AESPA activations~\cite{park2022aespa} on CIFAR-10 against EasyFHE.
For BERT-base, the baselines are EasyFHE's upstream THOR~\cite{moon2025thor} implementation (revision \texttt{4a2a2d6}) and a dedicated EasyFHE schedule for the MOAI encoder~\cite{zhang2026moai}.
Within each workload, the schemes use the same model weights, CKKS parameters, input samples, and output boundary.
We compare classification logits for ResNet-20 and THOR and final encoder hidden states for MOAI, using one author-provided input padded to 128 tokens for MOAI.
Operator benchmarks process equal numbers of real inputs.

\subsection{End-to-End Inference}
\label{sec:eval-e2e}
Table~\ref{tab:e2e} reports amortized online time per item, throughput, peak active GPU memory, and maximum normalized output error relative to plaintext inference.

\textbf{ResNet-20 with AESPA activations.}
\projectname{} processes two CIFAR-10 images in 651.14\,ms on the RTX 5090 and 655.01\,ms on the RTX PRO 6000, achieving $2.06\times$ and $2.05\times$ speedups over independent real execution and sustaining a $1.97\times$ speedup over the fastest native batched-squaring baseline on both GPUs.
\projectname{} packs both images into one ciphertext and keeps them packed through convolutions and AESPA square activations.
Across five refresh points, packing reduces the number of bootstraps from ten per-image calls to five dual-lane calls.
Section~\ref{sec:eval-ablation} separates the gain from sharing these refreshes from the gain from keeping both images packed between them.

On both GPUs, peak active GPU memory increases by less than $1\%$ over native batching (14.98 to 15.11\,GiB), while maximum normalized output error changes from $4.09\times10^{-6}$ to $5.53\times10^{-6}$ on RTX 5090 and from $4.19\times10^{-6}$ to $5.48\times10^{-6}$ on RTX PRO 6000.

\textbf{BERT-base inference.}
On THOR, \projectname{} preserves EasyFHE's attention layout and keeps feature pairs packed through refresh and GELU in each feed-forward network.
On the RTX 5090, median online latency after warmup falls from 52.077 to 39.256\,s over 408 MRPC validation pairs.
Peak active memory on this GPU includes allocations retained between the two schemes in the same process.
On the RTX PRO 6000, mean online latency falls from 43.211 to 35.973\,s over two fixed validation pairs, with resident model weights and a 48\,GiB weight-encoding cache in both schemes.

On the RTX PRO 6000, MOAI latency falls from 16.847 to 13.902\,s per position ($1.212\times$), including input upload and amortizing batch time over 256 positions.
Across the final hidden states, the maximum normalized output errors against the same plaintext reference are 0.353 and 0.00731, respectively.

\subsection{Nonlinear Circuit Evaluation}
\label{sec:eval-micro}
\textbf{Dual-lane squaring.}
On the RTX 5090, $\operatorname{DSq}$ is $1.02$--$1.14\times$ faster than two independent real squares across 48 tested configurations, including one conjugation key switch in the packed timing.
\begin{table}[ht]
\centering
\caption{Median nonlinear circuit latency (ms) on the RTX 5090.}
\label{tab:selected-replay}
\footnotesize
\setlength{\tabcolsep}{2.5pt}
\renewcommand{\arraystretch}{1.12}
\begin{tabular*}{\columnwidth}{@{\extracolsep{\fill}}lccccc@{}}
\toprule
\multicolumn{6}{c}{\textbf{(a) Polynomial activations}}\\
Function & \shortstack{Direct PS\\(real)} & \shortstack{Sym. PS\\(real)} & \shortstack{Sym. PS\\(dual)} & HMult count & Speedup\\
\midrule
GELU    & 6.269  & 5.847  & \textbf{3.902} & $24\to22\to16$ & $1.61\times$\\
Tanh    & 13.546 & 11.458 & \textbf{7.583} & $38\to32\to25$ & $1.79\times$\\
Sigmoid & 8.019  & 6.617  & \textbf{4.585} & $26\to24\to18$ & $1.75\times$\\
SiLU    & 6.323  & 5.896  & \textbf{3.940} & $24\to22\to16$ & $1.60\times$\\
\end{tabular*}\par\nointerlineskip
\begin{tabular*}{\columnwidth}{@{\extracolsep{\fill}}lcccc@{}}
\midrule
\multicolumn{5}{c}{\textbf{(b) \textsf{EvalMod} and full bootstrapping}}\\
Stage & Baseline & Ours (off) & Ours (on) & Speedup\\
\midrule
\textsf{EvalMod} & 15.983 & 13.174 & \textbf{12.876} & $1.238\times$\\
Full bootstrapping & 32.608 & 29.694 & \textbf{29.537} & $1.104\times$\\
\bottomrule
\end{tabular*}
\par\smallskip
\parbox{\columnwidth}{(a) Real evaluation uses two ciphertexts and dual-lane evaluation uses one.
Sym.\ PS uses fixed symmetric circuits without conjugate propagation.
HMult counts follow the latency column order.
Speedups use real Direct PS as the baseline.
(b) Both packed schedules cache conjugates, and the on/off label indicates conjugate propagation.
Speedups use separate-lane EasyFHE as the baseline.}
\end{table}
\textbf{Polynomial activations.}
Across four activations, symmetry reduction with dual-lane execution is $1.60$--$1.79\times$ faster than direct Paterson--Stockmeyer (PS) evaluation~\cite{paterson1973polynomials} on two real ciphertexts (Table~\ref{tab:selected-replay}(a)).
When both implementations use the same reconstructed circuit, dual-lane execution retains a $1.44$--$1.51\times$ speedup.

\textbf{\textsf{EvalMod} and full bootstrapping.}
Table~\ref{tab:selected-replay}(b) compares separate-lane evaluation with scaled complex packing for sparse ternary CKKS at $N=2^{16}$.
Packing reduces median \textsf{EvalMod} latency from 15.983 to 13.174\,ms and full bootstrapping latency from 32.608 to 29.694\,ms.
With conjugate propagation enabled, these latencies fall to 12.876 and 29.537\,ms, yielding paired speedups of $1.238\times$ and $1.104\times$ over the separate-lane baseline.

\subsection{Packing Across ResNet-20 Layers}
\label{sec:eval-ablation}
We compare packing only at refresh with keeping the two images packed throughout ResNet-20 on the RTX 5090.
Both packed schedules perform five bootstraps and spend about 311\,ms on refresh, including slot layout conversion.
Over 50 two-image replays, keeping the images packed through convolutions and square activations reduces median non-refresh time from 630.54 to 348.40\,ms and median end-to-end latency from 942.07 to 659.79\,ms, a $1.43\times$ speedup over refresh-only packing.

%% file: sections/related-work.tex
\section{Related Work}
\label{sec:related_work}

\noindent\textbf{Complex packing and nonlinear evaluation.}
Xu and Wang~\cite{xu2025multiimage} use complex slots for batched convolutions and separate the lanes for activations.
THOR~\cite{moon2025thor} and MOAI~\cite{zhang2026moai} optimize tensor layouts and nonlinear evaluation in encrypted Transformers, while FEnc$^2$~\cite{ran2026fenc2} reduces rotations and ciphertext occupancy in CNNs.
AutoFHE~\cite{ao2024autofhe} selects activation polynomial degrees and bootstrapping points; RBOOT~\cite{yang2026rboot} integrates ReLU evaluation into CKKS bootstrapping.
\projectname{} addresses lane coupling by sharing nonlinear squares and preserving independent lane outputs across compatible operators.

\noindent\textbf{FHE compilers.}
EVA~\cite{dathathri2020eva} manages ciphertext levels and scales; Fhelipe~\cite{krastev2024fhelipe} selects tensor layouts, Rotom~\cite{chen2026rotom} optimizes layout transitions, and HEIR~\cite{ali2025heir} offers reusable compiler abstractions.
EasyFHE~\cite{easyfhe2026} lowers encrypted programs with GPU and memory optimizations; Cerium~\cite{jayashankar2026cerium} supports large-scale multi-GPU CKKS execution, and Libra~\cite{bian2026libra} co-optimizes cross-scheme patterns and scheduling.
\projectname{} selects shared-square circuits under layout and ciphertext-state constraints and plans their GPU execution.

\noindent\textbf{GPU and hardware acceleration.}
TensorFHE and Cheddar optimize GPU arithmetic~\cite{fan2023tensorfhe,choi2026cheddar}; HyperDrive and Helios target memory traffic and kernel fusion, respectively~\cite{fan2026hyperdrive,chen2026helios}.
Hades automates FHE accelerator design~\cite{liu2026hades}, and Chimera supports both CKKS and TFHE in one accelerator~\cite{hua2026chimera}.
\projectname{} complements such acceleration by reducing the number of nonlinear operations during circuit construction, before memory-aware GPU scheduling and lowering.

%% file: sections/conclusion.tex
\section{Conclusion}
\label{sec:conclusion}

We presented \projectname{}, a GPU compiler that uses complex packing to share nonlinear computation across two real streams in CKKS inference.
Dual-lane squaring uses one HMult, and shared-square circuits keep the lanes packed through polynomial activations and \textsf{EvalMod}.
Ciphertext state alignment, conjugate reuse, and GPU memory planning carry this packing across compatible operators.
On both the RTX 5090 and the RTX PRO 6000, \projectname{} achieves a $1.97\times$ end-to-end speedup for two-image AESPA ResNet-20 inference over the fastest tested EasyFHE batched baseline.
For BERT-base inference, the speedups over EasyFHE's THOR implementation are $1.33\times$ and $1.20\times$, respectively.